\input harvmac.tex
\noblackbox

%%%%%%%%%%%%%%%%%%%%%%%%%%%%%%%%%%%%%%%%%%%%%%%%%%%%%%%%%%%%%%%%%%%%%%%%%%%
%Blackboard letters
%  The prehistoric version of this font is known as "msym". Many unfortunate
%  souls still have this old (and UGLY) ancestor of "msbm". Time to join the
%  modern world guys!

\font\blackboard=msbm10 \font\blackboards=msbm7
\font\blackboardss=msbm5
\newfam\black
\textfont\black=\blackboard
\scriptfont\black=\blackboards
\scriptscriptfont\black=\blackboardss
\def\blackb#1{{\fam\black\relax#1}}

%   Those truly poor slobs who have neither "msbm not "msym" fonts can
% substitute
%   the definition

%\def\blackb{\bf}

%   for the above font definitions or, if all else fails,
%   return to scratching symbols in the dirt with a sharpened stick.
%
\def\BC{{\blackb C}} 
 
\def\BZ{{\blackb Z}} 
\def\BP{{\blackb P}}

% Blackboard bold "1". Not in the AMS font set.

%%%%%%%%%%%%%%%%%%%%%%%%%%%%%%%%%%%%%%%%%%
% Math boldface letters
%
\font\mathbold=cmmib10 \font\mathbolds=cmmib7
\font\mathboldss=cmmib5
\newfam\mbold
\textfont\mbold=\mathbold
\scriptfont\mbold=\mathbolds
\scriptscriptfont\mbold=\mathboldss
\def\bi{\fam\mbold\relax}

\def\CO{{\cal O}}\def\CW{{\cal W}}
\def\H#1#2{{\rm H}^{#1}(#2)}
\def\cp#1{{\BC{\rm P}^{#1}}}
\def\wp#1{{W\BP^{#1}}}
\def\pd#1#2{{\partial #1\over\partial #2}}
%semi-direct product |><
\def\vev#1{\langle #1\rangle}
\def\ket#1{|#1\rangle}
\def\CE{{\cal E}}

\def\Ka{K\"ahler}
\def\cy{Calabi--Yau}
\def\LG{Landau--Ginzburg}
\def\sm{$\sigma$-model}
\def\lsm{linear $\sigma$-model}
\long\def\optional#1{}

\lref\DGExact{J. Distler and B. Greene,
``Some Exact Results on the Superpotential from Calabi--Yau
Compactifications,'' {\it Nucl. Phys.} {\bf B309} (1988) 295.}
\lref\Kawai{T. Kawai and K. Mohri, ``Geometry of (0,2) Landau--Ginzburg
Orbifolds,'' {\it Nucl. Phys.} {\bf B425} (1994) 191, {\tt hep-th/9402148}.}
\lref\Martinec{E. Martinec,
``Criticality, Catastrophes, and Compactifications,''
in {\it Physics and Mathematics of Strings}, ed. L. Brink,
D. Friedan, and A.M. Polyakov, World Scientific, 1992.}
\lref\Vafa{C. Vafa and N. Warner, ``Catastrophes and the Classification
of Conformal Field Theories,'' {\it Phys. Lett.} {\bf 218B}
(1989) 51; B.R. Greene, C. Vafa, and N.P. Warner, ``Calabi--Yau Manifolds
and Renormalization Group Flows,'' {\it Nucl. Phys.} {\bf B324}
(1989) 371.}
\lref\SW{A. Strominger and E. Witten, ``New Manifolds for
Superstring Compactification,'' {\it Comm. Math. Phys.}
{\bf 101} (1985) 341.}
\lref\duality{A. Giveon, M. Porrati, and E. Rabinovici, ``Target
Space Duality in String Theory,'' {\it Phys. Rept.} {\bf 244} (1994)
77, {\tt hep-th/9401139}.}
\lref\DK{J. Distler and S. Kachru, ``(0,2) Landau--Ginzburg Theory,''
{\it Nucl. Phys.} {\bf B413} (1994) 213, {\tt hep-th/9309110}.}
\lref\DKtwo{J. Distler and S. Kachru, ``Singlet Couplings and (0,2)
Models,'' {\it Nucl. Phys.} {\bf B430} (1994) 13, {\tt hep-th/9406090}.}
\lref\Witten{E. Witten, ``Phases of N=2 Theories in Two Dimensions,''
{\it Nucl. Phys.} {\bf B403} (1993) 159, {\tt hep-th/9301042}.}
\lref\NewIss{E. Witten, ``New Issues in Manifolds of SU(3) Holonomy,''
{\it Nucl. Phys.} {\bf B268} (1986) 79.}
\lref\Mirror{S.-T. Yau, ed., {\it Essays on Mirror Manifolds},
International Press, 1991.}
\lref\UY{K. Uhlenbeck and S.-T. Yau, ``On the Existence of
Hermitian--Yang--Mills Connections in Stable Vector Bundles,''
{\it Comm. Pure. App. Math.} Vol. XXXIX (1986) S257.}
\lref\WitSilvtwo{E. Silverstein and E. Witten, ``Criteria for Conformal
Invariance of (0,2) Models,'' {\it Nucl. Phys.} {\bf B444} (1995) 161,
{\tt hep-th/9503212}.}
\lref\Dixon{T. Banks, L. Dixon, D. Friedan, and E. Martinec,
``Phenomenology and Conformal Field Theory or Can String Theory Predict
the Weak Mixing Angle?,'' {\it Nucl. Phys.} {\bf B299} (1988) 613.}
\lref\CHSW{P. Candelas, G. Horowitz, A. Strominger, and E. Witten,
``Vacuum Configurations for Superstrings,'' {\it Nucl. Phys.} {\bf B258}
(1985) 46.}
\lref\KW{S. Kachru and E. Witten, ``Computing the Complete Massless
Spectrum of a Landau--Ginzburg Orbifold,'' {\it Nucl. Phys.}
{\bf B407} (1993) 637, {\tt hep-th/9307038}.}
\lref\DistGr{J. Distler and B. Greene, ``Aspects of (2,0) String
Compactifications,'' {\it Nucl. Phys.} {\bf B304} (1988) 1.}
\lref\DSWW{M. Dine, N. Seiberg, X. Wen,  and E. Witten,
``Non-Perturbative Effects on the String World Sheet I,II,'' {\it
Nucl. Phys.} {\bf B278} (1986) 769, {\bf B289} (1987) 319.}
\lref\rAGM{P.S. Aspinwall, B.R. Greene, and D.R. Morrison, ``Calabi--Yau
Moduli Space, Mirror Manifolds, and Spacetime Topology Change in
String Theory,'' {\it Nucl. Phys.} {\bf B416} (1994) 414,
{\tt hep-th/9309097}.}
\lref\rGP{B.R. Greene and M.R. Plesser, ``Duality in Calabi--Yau Moduli
Space,''
{\it Nucl. Phys.} {\bf B338} (1990) 15.}
\lref\CP{P. Candelas, X. de la Ossa, P. Green, and L. Parkes, ``A Pair
of Calabi--Yau Manifolds as an Exactly Soluble Superconformal Theory,''
{\it Nucl. Phys.} {\bf B359} (1991) 21.}
\lref\models{S. Kachru, ``Some Three Generation (0,2) Calabi--Yau Models,''
Harvard preprint to appear.}
\lref\Candelas{P. Candelas, ``Yukawa Couplings Between (2,1) Forms,''
{\it Nucl. Phys.} {\bf B298} (1988) 458.}
\lref\rAG{P.S. Aspinwall and B.R. Greene,
``On the Geometric Interpretation of N = 2 Superconformal Theories,''
{\it Nucl. Phys.} {\bf B437} (1995) 205, {\tt hep-th/9409110}.}
\lref\MP{D.R. Morrison and M.R. Plesser,
``Summing the Instantons: Quantum Cohomology and Mirror Symmetry in Toric
Varieties,'' {\it Nucl. Phys.} {\bf B440} (1995) 279, {\tt hep-th/9412236}.}
\lref\rBat{V.V. Batyrev, ``Dual Polyhedra and Mirror Symmetry for Calabi--Yau
Hypersurfaces in Toric Varieties,'' {\it J. Algebraic Geometry} {\bf 3}
(1994) 493, {\tt alg-geom/9310003}.}
\lref\BatBor{V.V. Batyrev and L.A. Borisov,
``Dual Cones and Mirror Symmetry for Generalized {C}alabi--{Y}au
  Manifolds,''
To appear in: {\it Essays on Mirror Manifolds II}, {\tt alg-geom/9402002}.}
\lref\BatQcoh{V.V. Batyrev, ``Quantum Cohomology Rings of Toric
Manifolds,'' {\it Journ\'ees de
  G\'eom\'etrie Alg\'ebrique d'Orsay (Juillet 1992)}, Ast\'erisque, vol. 218,
  Soci\'et\'e Math\-\'ematique de France, 1993, p.~9, {\tt alg-geom/9310004}.}
\lref\CdK{P. Candelas, X. de la Ossa, and S. Katz, ``Mirror Symmetry for
Calabi--Yau Hypersurfaces in Weighted $P_4$ and
Extensions of Landau Ginzburg Theory,'' {\it  Nucl. Phys.} {\bf B450}
(1995) 267, {\tt hep-th/9412117}.}
\lref\rGMS{B.R. Greene, D.R. Morrison, and A. Strominger,
``Black Hole Condensation and the Unification of String Vacua,''
{\it Nucl. Phys.} {\bf B451} (1995) 109, {\tt hep-th/9504145}.}
\lref\CdFKM{P. Candelas, X. de la Ossa, A. Font, S. Katz, and
D.R. Morrison, ``Mirror Symmetry for Two Parameter Models -- I,''
{\it Nucl. Phys.} {\bf B416} (1994) 481, {\tt hep-th/9308083}.}
\lref\HKTY{S. Hosono, A. Klemm, S. Theisen, and S.-T. Yau,
``Mirror Symmetry, Mirror Map and Applications to Calabi--Yau
Hypersurfaces,'' {\it Commun. Math. Phys.} {\bf 167} (1995) 301,
{\tt hep-th/9308122}.}
\lref\BSW{R. Blumenhagen, R. Schimmrigk, and A. Wisskirchen,
``The (0,2) Exactly Solvable Structure of Chiral Rings,
Landau--Ginzburg Theories and Calabi--Yau Manifolds,''
{\it Nucl. Phys.} {\bf B461} (1996) 460, {\tt hep-th/9510055}.}
\lref\WitSilv{E. Silverstein and E. Witten,
``Global U(1) R-Symmetry And Conformal Invariance Of (0,2) Models,''
{\it Phys. Lett.} {\bf B328} (1994) 307, {\tt hep-th/9403054}.}
\lref\DKthree{J. Distler and S. Kachru, ``Duality of (0,2) String Vacua,''
{\it Nucl. Phys.} {\bf B442} (1995) 64, {\tt hep-th/9501111}.}
\lref\DNotes{J. Distler, ``Notes on (0,2) Superconformal Field Theories,"
Proceedings of the 1994 Trieste Summer School, {\tt hep-th/9502012}.}
\lref\BSV{M. Bershadsky, V. Sadov, and C. Vafa, ``D-Strings on
D-Manifolds,'' {\tt hep-th/9510225}.}
\lref\Strominger{A. Strominger, ``Massless Black Holes and Conifolds in
String Theory,'' {\it Nucl. Phys.} {\bf B451} (1995) 96, {\tt hep-th/9504090}.}
\lref\Aspinwall{P.S. Aspinwall, ``Enhanced Gauge Symmetries and K3
Surfaces,'' {\it Phys. Lett.} {\bf B357} (1995) 329, {\tt hep-th/9507012}.}
\lref\dynamics{E. Witten, ``String Theory Dynamics In Various Dimensions,''
{\it Nucl. Phys.} {\bf B443} (1995) 85, {\tt hep-th/9503124}.}
\lref\comments{E. Witten, ``Some Comments On String Dynamics,''
{\tt hep-th/9507121}.}
\lref\smallinstantons{E. Witten,
``Small Instantons in String Theory,'' {\tt  hep-th/9511030}.}
\lref\DMW{M.J. Duff, R. Minasian, and E. Witten, ``Evidence for
Heterotic/Heterotic Duality'', {\tt hep-th/9601036}.}
\lref\rKSS{S. Kachru, N. Seiberg and E. Silverstein, ``SUSY Gauge Dynamics
and Singularities of $4d$ N=1 String Vacua'', {\tt hep-th/9605036}.}

\Title{\vbox{\hbox{CLNS 96/1411}\hbox{DUKE-TH-96-109}\hbox{UTTG--05--96}
\hbox{\tt hep-th/9605222}\vskip -.5in }}
{Resolving Singularities in (0,2) Models}
\centerline{Jacques Distler$^\star$, Brian R. Greene$^\dagger$ and David R.
Morrison$^\ddagger$}
%\bigskip
\vskip.3in
\centerline{\hbox{
\vtop{\hsize=1in\it
\centerline{$^\star$Theory Group}
\centerline{Department of Physics}
\centerline{University of Texas}
\centerline{Austin, TX 78712}
}\hskip .125in
\vtop{\hsize=3in\it
\centerline{$^\dagger$F.R. Newman Laboratory}
\centerline{of Nuclear Studies}
\centerline{Cornell University}
\centerline{Ithaca, NY 14853}
}\hskip .125in
\vtop{\hsize=1.25in\it
\centerline{$^\ddagger$Department of Mathematics}
\centerline{Box 90320}
\centerline{Duke University}
\centerline{Durham, NC 27708}
}
}}

{\parindent=-5pt

\footnote{}{\par
${}^\star$Email: {\tt distler@golem.ph.utexas.edu}\ .\par
${}^\dagger$Email: {\tt greene@hepth.cornell.edu}\ .\par
${}^\ddagger$Email: {\tt drm@math.duke.edu}\ .\par
}
}

%\vskip .25in
\vskip.5in

In contrast to the familiar $(2,2)$ case, the singularities which
arise in the $(0,2)$ setting can be associated with degeneration of
the base Calabi--Yau manifold {\it and/or}\/ with degenerations of the
gauge bundle. We study a variety of such singularities and give a
procedure for resolving those which can be cured perturbatively.
Among the novel features which emerge are models in which smoothing
singularities in the base yields a gauge {\it sheaf}\/ as opposed to a
gauge bundle as the structure to which left moving fermions
couple. Supersymmetric $\sigma$-models with target data being an
appropriate sheaf on a Calabi--Yau space therefore appear to be the
natural arena for $N = 1$ string models in four dimensions.  We also
indicate a variety of singularities which would require a
nonperturbative treatment for their resolution and briefly discuss
applications to heterotic models on K3.

\bigskip

\Date{\it May 1996} %replace this line by \draft  for preliminary versions

%\draft

\newsec{Introduction}

After many years of intensive
investigation, string compactifications with $(2,2)$ world sheet supersymmetry
continue to yield new and remarkable physical consequences.
It is important to  realize, though, that $(2,2)$ models are
likely to be but
a small slice through the more general class of $(0,2)$ compactifications.
Historically, $(0,2)$ Calabi--Yau compactifications have received less
attention
because they are technically more difficult to construct and to analyze than
their
$(2,2)$ counterparts. The work of \Witten\ went a long way
towards ameliorating this unpleasant aspect by providing a new tool---the
linear $\sigma$-model---for dealing with both $(2,2)$ and $(0,2)$ models.
The linear $\sigma$-model provides a non-conformal member of the universality
class
of a superconformal theory which captures many features of the latter while
avoiding much of its complexity. Furthermore, the linear $\sigma$-model
provides
a bridge between (non-conformal) Calabi--Yau $\sigma$-models, with either
$(2,2)$
or $(0,2)$ world sheet supersymmetry, and Landau--Ginsburg mean field theory
models. The latter are well understood, relatively easy to analyze and share
a number of important physical characteristic with the Calabi--Yau's to which
they are connected. Hence, they provide another important tool for detailed
study. A number of papers
\refs{\DK\WitSilv\DKtwo\DKthree\WitSilvtwo{--}\BSW}
have used these new tools to
initiate a comprehensive investigation of $(0,2)$ models. It is
important to note that in
\refs{\Witten,\rAGM}
and in greater detail in \refs{\rAG,\MP} it was shown that
methods of toric geometry are equivalent to those of the linear $\sigma$-model
but in certain circumstances provide a more powerful analytic tool.
We shall avail ourselves of this approach  in the sequel.

Ultimately we hope to have as complete an understanding of $(0,2)$ models
as we presently have for $(2,2)$ models. Although this goal is still rather
far off, the work of \refs{\DK\WitSilv\DKtwo\DKthree\WitSilvtwo{--}\BSW}
and, hopefully, the present paper, are
steps in this direction. More specifically, the mathematics and physics of
apparent singularities in a variety of contexts has played a key role
in numerous recent developments in string theory and in field theory.
This  is true, in particular, for $(2,2)$ string compactifications.
The mirror symmetry construction of \rGP, for example, relies on Calabi--Yau
orbifolds
which generally have singularities.
The phase structure of $(2,2)$ moduli space found in \refs{\Witten,\rAGM}
shows that parameter spaces for numerous conformal theories adjoin along
common
walls, which are geometrically interpretable as singular configurations.
This phenomenon was dramatically augmented through the work of
\refs{\Strominger,\rGMS} in which, at the level of nonperturbative type II
string
theory,
many and possibly all vacuum configurations were shown to join together
through
mathematically singular but physically smooth transitions. And much of the
exciting
work on string dualities
focuses on various singularities
as key points of physical interest
\refs{\dynamics\Aspinwall\comments{--}\BSV}. It therefore seems quite important
to
understand both the mathematics and the physics of singularities
in $(0,2)$ Calabi--Yau
conformal field theory. In this paper we begin such a study.

In section II we review the linear $\sigma$-model/toric geometry approach
to $(2,2)$ models and show how it naturally fits singular configurations into
a phase diagram which contains appropriate desingularizations. We then
review the linear $\sigma$-model approach to $(0,2)$ models, as presented in
\refs{\Witten,\DK,\DNotes}, and indicate the singularities which arise.
In section III we give a procedure for extending the  phases analysis to
the $(0,2)$ case and thereby resolving the singularities encountered.
Although adequate for resolving singularities we still seek a more unified
toric treatment.
In the course of resolving $(0,2)$ singularities, we
shall find a number of interesting physical differences from the
$(2,2)$ case.
After pointing out these differences, we illustrate them
in section IV with a number
of explicit examples. In section V we give some brief conclusions and
indicate directions for future work.

\newsec{$(2,2)$ and $(0,2)$ Models: A Linear $\sigma$-Model Approach}

\subsec{Bosonic Fields}

We begin with a brief discussion of the linear $\sigma$-model introduced in
\Witten, and its extension discussed
in \rAG\ to which the reader should refer for more detail.
Rather than being completely general, we review the
case which corresponds to a Calabi--Yau hypersurface in   a weighted
projective four space. Later in this paper we will consider some
generalizations.

Witten found that an interesting class of two dimensional models with
$(2,2)$ world sheet supersymmetry could be constructed
by starting with  an $N = 2$ supersymmetric gauge theory
 with gauge group $U(1)$ and action
\eqn\eactionw{ S = S_{\rm  kinetic} + S_{W} + S_{\rm gauge} + S_{\rm FI-D\
term} .}
The term $S_W$ takes the form
\eqn\eSW{ S_W = \int d^2z d^2 \theta W(P,S_1,...,S_5) }
where $W$ is the superpotential of the theory, $P,S_1,...,S_5$ are chiral
superfields whose $U(1)$ charges are denoted $q_0,q_1,..,q_5$
and $W$ is chosen to be a
$U(1)$ invariant holomorphic function of the form
\eqn\eWW{ W = P G(S_1,...,S_5).}
In this expression $G$ is a transverse quasihomogeneous function of
$S_1,...,S_5$
whose overall $U(1)$ charge is $-q_0$
For future reference we note that we can explicitly integrate out
one of the superspace coordinates in \eSW, say $\theta^{-}$ and write this
contribution
to the action as
\eqn\eSWzt{\int d^2z d\theta \Gamma G + P \Lambda_i F_i }\foot{
We use $\theta^{+}$ and $\overline \theta^{+}$ as our
right moving (i.e. $(0,2)$) fermionic coordinates and
$\theta^{-}$ and $\overline \theta^{-}$
as our left moving ($(2,0)$) fermionic coordinates. Hence $+,-$ refer to
right moving and left moving, respectively, and  an overline denotes opposite
$U(1)$ eigenvalue.}.
 In
this expression
we have expressed a general $(2,2)$ chiral superfield in
terms of its $(0,2)$ chiral field content, namely,
\eqn\edecomp{
\Phi^{(2,2)} = \Phi^{(0,2)} + \theta^{-} \Psi^{(0,2)} + i\theta^-
\overline\theta^-(-\del
\Phi^{(0,2)})}
 where $\Phi^{(0,2)}$ and  $\Psi^{(0,2)}$ are $(0,2)$ bosonic
and fermionic multiplets, respectively. Our notation is that, in
the $(2,2)$ context, $P^{(0,2)}$ and $\Gamma^{(0,2)}$ constitute
$P^{(2,2)}$, while $S_i^{(0,2)}$ and  $\Lambda_i^{(0,2)}$
constitute $\Phi_i^{(2,2)}$. The quantity $F_i$ denotes $\partial G \over
\partial
S_i$.
Typically we will drop the
$(2,2)$ and $(0,2)$ superscripts, as we have done in all previous equations.
The
Fayet--Illiopoulos $D$-term takes the form
\eqn\eFID{S_{\rm FI-D\ term} = t \int d \theta^+ d \overline \theta^-
\Sigma + c.c.}
where $\Sigma$ is a twisted chiral superfield and $t = r + i \theta$ is
a complex parameter.
Witten showed that this model has a nontrivial phase structure in the sense
that for $r$ large and positive it reduces in the infrared to a Calabi--Yau
$\sigma$-model on the Calabi--Yau space $G = 0$ in $\wp{4}_{q_1,q_2,...,q_5}$
with homogeneous coordinates $(s_1,...,s_5)$
($s_i$ is the scalar part of the superfield $S_i$) while
for $r$ large and negative it reduces to a Landau--Ginsburg model with
superpotential
given by $G$. Seeing this is a straightforward exercise in studying the
bosonic
potential
\eqn\ebosonic{U = |G(s_i)|^2 + |p|^2 \sum_i |{{\partial G \over \partial
s_i}}|^2 +
{{1 \over 2 e^2}} D^2 + 2|\sigma|^2 ( \sum_i q_i^2|s_i|^2 + q_0^2 |p|^2) }
with
\eqn\eD{ D = -e^2( \sum_i q_i |s_i|^2 - q_0 |p|^2 - r)} for the two cases
distinguished
by the sign of $r$. For
$ r > 0,$ vanishing of the potential
requires $p = 0, \sigma = 0, G = 0, \sum_i q_i |s_i|^2 = r$. Together
with the $U(1)$ gauge symmetry identifications, this yields the
stated Calabi--Yau $\sigma$-model. For $ r < 0$, we find $s_i = 0,
\sigma = 0, p = \sqrt(-r/q_0)$ and the resulting model describing
fluctuations around this vacuum configuration is a Landau--Ginsburg model
with potential proportional to $G$. These two descriptions of the physical
model can be thought of as different phases of the overarching
Lagrangian \eactionw.
The fact that $r$ is actually part of a complex parameter which
includes a theta angle $t = r + i \theta$  plays a key role in
establishing that the transition from one phase to the other is smooth
\Witten.

An important point for the present study is that the typical Calabi--Yau
obtained in this manner is singular as it is embedded in the
weighted projective space $\wp{4}_{q_1,q_2,...,q_5}$ which itself has
singularities unless all of the charges are relatively prime.
The physical model is well behaved even with these singularities,
but their presence signals that the phase analysis indicated above,
for such a model, is  incomplete. Namely, there are marginal operators
associated with resolving the singularities. These marginal
operators can be used to deform the original model to a desingularized
form and thereby probe regions
of the moduli space not encountered by our previous discussion.
In particular, in the above discussion  we assumed that we had
a single $U(1)$ gauge symmetry which gave rise to one K\"ahler moduli
space parameter $r$. As indicated, this is but a one-dimensional slice through
the
complete moduli space and hence we are led to embed this linear $\sigma$-model
in one which has a $U(1)^s$ gauge symmetry with corresponding parameters
$r_1,...,r_s$ where $s$ equals $h^{11}$ of the resolved space.
To do so, we need to know the charges of our fields under this full
gauge symmetry group. Furthermore, each $U(1)$ factor gives rise to its
own Fayet--Illiopoulos $D$-term whose vanishing cuts down the vacuum
configuration
by one complex dimension. Since this dimension is fixed, each of the $s - 1$
additional
$U(1)$ factors must be accompanied by an additional chiral superfield and
hence
we also need to know the $U(1)^s$ charges of
these additional $s - 1$ chiral superfields $\chi_1,...,\chi_{s-1}$.
Knowledge of this data provides us with a moduli space containing that of
the original singular Calabi--Yau, but enlarged to include regions
corresponding
to its desingularization. How, therefore, do we find this data? Methods
of toric geometry prove to be the most efficient and systematic way of
doing so.

The link between the linear $\sigma$-model and toric geometry arises because
mathematically, setting the $D$-term to zero and taking proper account
of the $U(1)$ phase symmetry corresponds to taking a {\it symplectic}\/
quotient. As is well known, this can  be rephrased as a holomorphic quotient
and toric geometry is a formalism for studying the latter.
%\ref\r{References}.
Roughly speaking,
toric geometry provides a systematic method for studying spaces that
can be realized as holomorphic quotients of the form
$(\BC^n - F_{\Delta})/(\BC^*)^m$ with $F_{\Delta} \subset \BC^n$.
A weighted projective space such as
$\wp{4}_{q_1,q_2,...,q_5}$, for example, can be realized in this manner via
$(\BC^5 - (0,0,0,0,0))/\BC^*$ with $\BC^*$ action being
$(z_1,...,z_5) \rightarrow (\lambda^{q_1} z_1,..., \lambda^{q_5} z_5)$,
$\lambda \in \BC^*$.
Toric geometry gives us a simple procedure for desingularizing
such holomorphic quotients which we now briefly review.
The reader should consult
\refs{\rAGM,\rAG,\MP,\CdK}
for more details.

As in \refs{\BatBor,\rAGM} the toric varieties of interest to us
are associated to
reflexive
Gorenstein cone in $\BC^5$ with apex at the origin and edges given by
five vectors $v_1,...,v_5$.
Phases of the physical model correspond to distinct triangulations of
a transverse hyperplane section of the cone lying at a unit distance from
the origin---that is, triangulations of the polytope with vertices given
by the above edges. To be concrete, consider the example of the quintic
hypersurface
with $(q_0,...,q_5) = (-5,1,1,1,1,1)$ and edges
\eqn\equintic{\vbox{\centerline{$v_1=(1,0,0,0,1),
v_2=(0,1,0,0,1), v_3=(0,0,1,0,1),$}\centerline{$v_4=(0,0,0,1,1),
v_5=(-1,-1,-1,-1,1).$}}}
We see that there are two possible triangulations of
this polytope: the
triangulation consisting of the  polytope itself and the triangulation
of the polytope into the five sections with vertices
$\{v_0,...,\hat v_i,...,v_5\}$ where $v_0$ is the interior
point $(0,0,0,0,1)$ and $\hat v_i$ denotes omitting the $i^{th}$ vertex.
Toric geometry associates each of these triangulations to holomorphic
quotients of the the form $(\BC^6 - F_{\Delta})/\BC^*$ where the
form of $F_{\Delta}$ is determined by the triangulation
and the action of $\BC^*$ is determined by the point set $v_0,...,v_5$,
as explained in \refs{\rAGM,\rAG}. In this example
with $\BC^6$ variables labeled $(s_1,...,s_5,p)$, the first triangulation
yields
\eqn\eFDELTAONE{ F^{(1)}_{\Delta} = (s_1,...,s_5,0) }
while the second gives
\eqn\eFDELTATWO{ F^{(2)}_{\Delta} = (0,...,0,p) }
both with $\BC^*$ action
\eqn\eCSTAR{(s_1,...,s_5,p) \rightarrow (\lambda s_1,...,\lambda
s_5,\lambda^{-5} p).}
Examination of these holomorphic quotients reveals that the first
corresponds to the $\BC^5/\BZ_5$  configuration space of the $r < 0$ phase
while the
second corresponds to the
${\cal O}(-5)$ over $\cp{4}$ $r > 0$ phase of the linear $\sigma$-model.
The quintic itself is recovered in the $r >0$ phase by requiring that the full
bosonic
potential vanishes thereby yielding the locus of a quintic in $\cp{4}$.
In this way we see that the linear $\sigma$-model provides the physical
counterpart of such constructions of toric varieties.

We can make powerful use of this link since in more complicated examples
than the quintic, for which
the linear $\sigma$-model analysis becomes increasingly difficult, the
toric methods remain tractable. In particular, how do we resolve
the singularities which such toric constructions may yield?
 For a detailed
discussion see \refs{\rAGM,\rAG}; here we will only briefly summarize the
procedure.
Intuitively speaking, we want to excise a neighborhood of the singular loci
and glue in a smooth space which has the same boundary. The size and shape of
of the space we glue in are extra degrees of freedom that arise on the smooth
model.  As discussed earlier,
this  translates into the linear $\sigma$-model language as the existence
of more $U(1)$ gauge symmetries and more chiral superfields
giving a $U(1)^s$ gauge symmetry and chiral superfields
$P,S_1,...,S_5,\chi_1,...,\chi_{s-1}$.
As mentioned,
specifying the model requires that we give the $U(1)^s$ charges of all of
these
chiral superfields.
These charges are most systematically determined by the {\it kernel}\/ of the
toric
$(s + 5) \times 5$ point set matrix ${\cal A}$, describing
the polytopic base ${\cal P}$
of the associated reflexive Gorenstein cone as discussed in \rAG.
This, therefore, is how toric geometry gives us the linear $\sigma$-model
data for the resolved model.

In particular,
the linear $\sigma$-model in this augmented form has $s$ Fayet--Illiopoulos
$D$-terms
with coefficients $r_1,...,r_s$. The space of all possible values of these
parameters naturally divides up into phase regions in a manner similar to
the simple case of the quintic described above. One can determine these phase
regions by varying the  $r_1,...,r_s$ in the bosonic potential of the linear
$\sigma$-model and studying the result minima or, alternatively, by studying
the possible triangulations of ${\cal P}$. Often the latter approach is much
easier. Among the different phases are those which correspond to
Calabi--Yau $\sigma$-models on the possible (crepant) desingularizations of the
initial
Calabi--Yau space. These
correspond to maximal triangulations of  ${\cal P}$ \foot{
An important fact \rAGM\ is that there need not be a unique maximal
triangulation
and therefore there can be {\it distinct}\/ ways of resolving the Calabi--Yau
singularities (as the triangulation determines the set $F_{\Delta}$, their
difference arises in the identity of the latter). }
Thus, this linear $\sigma$-model/toric geometrical approach
provides a systemic procedure for resolving local quotient singularities and
clearly delineates the relationship between the various smooth and
singular geometric configurations.

All of the above discussion is in the context of $(2,2)$ models. Our interest
in this paper is in the larger class of $(0,2)$ models. The essential
difference between the two cases is the treatment of the left moving
world sheet fermions $\lambda^a$  which
in $(2,2)$ models
lie in supermultiplets with the world sheet scalars $\phi_i$
but do not in the $(0,2)$ case. That is,
from \edecomp\ we recall that a $(2,2)$ chiral
superfield decomposes into a $(0,2)$ bosonic
chiral superfield and a $(0,2)$ fermionic chiral
superfield, the latter containing
left moving chiral fermions, $\lambda$.
In the $(2,2)$ setting the properties of these fermions are determined
via the left moving supersymmetry from the properties of their bosonic
partners. In the $(0,2)$ case, though, the absence of the left moving
supersymmetry yields newfound independence for the left moving fermions
both in terms of their number and their interactions. To understand
the range of possibilities associated with these degrees of freedom, let
us first recall their properties in $(2,2)$ models and then pass to
the more general $(0,2)$ setting.

\subsec{Fermionic Fields}

We carry out our discussion in a fully resolved Calabi--Yau region of the
moduli space.
The $(2,2)$ world sheet supersymmetry implies that a complex scalar field
$\phi_i$ lies in a supermultiplet with both a left-moving $\lambda^i$ and a
right moving $\psi^i$ world sheet fermion.
Our goal is to study these theories in the far infrared and therefore
we need to determine which fermions are massless and hence survive
into the long distance limit.
 There are two types of terms
in the linear $\sigma$-model action which can give mass to these fermions.
The first comes from the gauge field part of the action and yields
couplings of the form
$\sum_i q_i^{(k)}\alpha_k \psi^i \phi_i + q_i^{(k)} \beta_k \lambda^i \phi_i$
where
$\alpha_k$ and $\beta_k$ are world sheet fermions from the $k^{th}$ vector
$U(1)$ vector multiplet, k = 1,...,s.
The second comes from the superpotential and yields couplings of the
form
$\sum_i (\gamma \psi^i + \pi \lambda^i) {{ \partial G \over \partial \phi_i}}$
where $\gamma$ and $\pi$ are the left and right moving fermionic
partners to $p$.
These couplings imply that the fermions which remain massless are those which
are in the kernel of the map
\eqn\emapg{
g: (\psi_1,..,\psi_5) \rightarrow \sum_i \psi^i {{ \partial G \over \partial
\phi_i}} }
(so that the second type of couplings do not contribute to their mass) but not
in the image of
\eqn\emapf{ f: (y_1,...,y_s) \rightarrow (\sum_k \, q_1^{(k)}y_k \phi_1,
 \sum_k \, q_2^{(k)}y_k \phi_2,...,\sum_k \, q_1 ^{(k)}y_k \phi_1)}
(so that the first type of couplings do not contribute to their mass).
Geometrically, we can interpret $\phi_i$ as a section of the bundle
${\cal O}(q_i^{(1)},q_i^{(2)},...,q_i^{(k)})$ over the resolved Calabi--Yau.
This is the line bundle whose first Chern class is
$\sum_k q_i^{(k)} J_k$ where the $J_k$ are $(1,1)$ forms generating the
integral
$2$-forms on the desingularized Calabi--Yau manifold.
Using this structure, we can describe the massless fermions as the cohomology
of the
(non-exact) sequence
\eqn\esequence{0 \rightarrow \oplus^k {\cal O} \rightarrow \oplus_{i=1}^6
{\cal O}(q_i^{(1)},q_i^{(2)},...,q_i^{(k)}) \rightarrow
 {\cal O}(q_0^{(1)},q_0^{(2)},...,q_0^{(k)}) \rightarrow 0.}
This same discussion, due to the left/right symmetry holds identically
for the fermions $\lambda^i$.

As previously
emphasized, we carried out this discussion in a smooth Calabi--Yau phase.
It is easily repeated in any other phase and hence we can  follow the smooth
tangent bundle on the resolved Calabi--Yau to any other phase, for instance,
to the phase associated with the original singular weighted projective space.
It is not hard to show that in this phase the massless fermions arise
from the cohomology of the sequence
\eqn\esingularseq{ 0 \rightarrow {\cal O} \rightarrow
 \oplus_{i=1}^5 {\cal O}(q_i) \rightarrow
{\cal O}(\sum_{i=1}^5 w_i) \rightarrow 0}
over  $\wp{4}_{q_1,...,q_5}$
as one would expect from a na\"{\i}ve application of the linear $\sigma$-model
ignoring all issues associated with singularities.

We can now see how the $(0,2)$ case differs from the $(2,2)$ case. In the
$(0,2)$ case the above discussion applies to the $\psi^i$'s as they are
still the superpartners to the $\phi_i$'s and hence have the same gauge
charges. However, the $\lambda$'s now have no relation to the $\phi_i$'s: both
the number of $\lambda$'s and their charges are data which are ingredients
in the  {\it definition}\/ of a $(0,2)$ model. This data must be chosen
in a consistent manner, i.e. to ensure anomaly freedom, but is otherwise
unconstrained. As we will discuss in more detail in the next section,
after such a choice has been made, we can re-run through the above discussion,
in
any phase and analyze the structure defining the massless fermions.
In the phase associated to the original singular weighted projective space,
the gauge bundle of the left moving fermions will take the form studied
in previous works such as \refs{\Witten,\DK,\DNotes}. Now, though, the
structure
of toric
geometry allows us to move from this phase to others such as smooth
Calabi--Yau phases. A consistent choice of gauge charges is geometrically
translated, in such a phase, into the data defining the resolution of
the gauge bundle. In the next section we describe this in greater detail.

\newsec{$(0,2)$ Singularities and Their Resolution}

As follows from the above discussion,
our $(0,2)$ models differ from $(2,2)$ models in
two essential ways:

1. The bosonic and fermionic chiral multiplets, which in a $(2,2)$ model
occur in pairs that join into a chiral multiplet, are independent both
in terms of their number and gauge charges.

2. The supersymmetric gauge transformations come in two
varieties---bosonic and fermionic gauge symmetries. Again, in
a $(2,2)$ model these occur in pairs which join into a chiral
multiplet parameterizing gauge transformations. In a $(0,2)$ model, though,
they are independent both in number and in their action on the fields
in the theory.

It is the above data---the list of bosonic and fermionic
$(0,2)$ chiral superfields along with the action of the bosonic
and fermionic gauge symmetries upon them, together with a gauge
invariant superpotential---which constitutes a $(0,2)$ model.
{}From the above discussion, it is clear that there are two complementary
ways of framing our discussion. We can describe this data, which is
necessary to construct a full phase diagram for a $(0,2)$ model, and
show how  in various phases it contains the singular
models of \refs{\Witten,\DK,\DNotes} while
in others these singularities are resolved. Or,
we can
start with the kind of models studied in \refs{\Witten,\DK,\DNotes} and show
how
to embed them in a larger phase diagram that includes fully resolved regions.
Lets take the latter approach.

For concreteness, we again work with base spaces that arise from
hypersurfaces in a weighted projective four space or desingularizations
thereof. From \refs{\Witten,\DK,\DNotes} and our discussion above, the
data we begin with
are the $U(1)$ gauge charges $q_1,...,q_5$ of the $(0,2)$ chiral
superfields $S_1,...,S_5$ (which contain
 $(s_1, \psi_1)$,...,$(s_1, \psi_1)$ as components)
\def\q{\tilde q}
and the charge $\q_0$ of a $(0,2)$ Fermi multiplet $\Gamma$ (which
contains $\gamma$ as a component).
We then seek a consistent choice of $U(1)$ gauge charges
$(\q_1,...,\q_n)$ of the $(0,2)$ Fermi multiplets $\Lambda_1,...,\Lambda_n$
(which contain $\lambda_1,...,\lambda_n$ as components) and also
the gauge charge $q_0$ of the $(0,2)$ chiral multiplet $P$ (which contains
$(p,\pi)$ as components).
The $(0,2)$ superpotential describing this model takes the form
\eqn\esuperpot{\int d^2z d\theta (\Gamma G + \Lambda^a P F_a)}
where $F_a$ are homogeneous polynomials in the chiral bosonic superfields
with $U(1)$ charges $-q_0 - \q_a$.
The reader may find it
instructive to compare this with \eSWzt\ where one sees
that the $F_a$ are  no longer determined by $G$ but rather are independent
degrees of freedom. We shall sometimes refer to the first term
in \esuperpot\ as $\Lambda^0F_0$ for uniformity of notation.
We see from this expression that we must choose
$\q_0 = -$(degree of homogeneity, $d$, of $G$) and
$q_0 = -$(degree of homogeneity of $F_a$ + $\q_a$) which must be independent
of $a$. Consideration of anomalies constrains these choices in the following
way:
\eqn\econe{\q_0 = d = \sum_{i=1}^5 q_i\ {\rm and}\ q_0 = -\sum_{a=1}^n \q_a}
\eqn\ectwo{\sum_{i = 0}^5 q_i^2 = \sum_{a=0}^n \q_a^2.}
There is  one immediate solution to these equations: $n = 5$ and
$\q_i = q_i$. This, of course, takes us back to a $(2,2)$ model.
More generally, though,  solutions of these equations (modulo
some other more subtle anomalies discussed a bit later) gives us
data defining a consistent $(0,2)$ model.

By studying the structure of the bosonic potential, one again finds that
this linear $\sigma$-model has two phases depending upon the sign of the
Fayet--Illiopoulos parameter $r$. For $r$ positive, the theory reduces, in the
infrared, to  a $(0,2)$ Calabi--Yau $\sigma$-model with base manifold given
by the vanishing of $G$ in $\wp{4}{q_1,...,q_5}$ (with right moving
fermions coupling to the tangent bundle of this Calabi--Yau \foot{More
precisely, we should
refer to the tangent bundle as a tangent V-bundle, since the Calabi-Yau
may have orbifold singularities and hence be a V-manifold. We will
typically not make this linguistic distinction.})
and left moving fermions
coupling to the bundle defined by the cohomology of the sequence
\eqn\esingularseqt{ 0 \rightarrow {\cal O} \rightarrow
 \oplus_{i=1}^n {\cal O}(\q_i) \rightarrow {\cal O}(\sum_{i=1}^n \q_i)
\rightarrow 0.}

As in the $(2,2)$ case,  singularities in the Calabi--Yau indicate
that we have not probed the full moduli space to which this model
belongs. In the $(2,2)$ setting, though, resolving
the Calabi--Yau space automatically resolved the tangent bundle and hence
determined how the left and right moving fermions behave in the desingularized
model. Concretely, the charges of the bosonic fields under the
full $U(1)^s$ gauge symmetry determines the charges
of the fermions as they are joined together
into $(2,2)$  supermultiplets. In the $(0,2)$
setting this is still true for the right moving fermions $\psi^i$ but it
is not true for the left moving fermions $\lambda^a$. Rather, we have to
resolve the Calabi--Yau space and then consider all possible ways of
consistently pulling back the gauge bundle to this smooth space.
Consistency here, from the geometrical point of view, is that the first
Chern class of the resolved bundle must vanish and its second Chern class
must equal that of the resolved tangent bundle. From a physical point of view,
this translates into
the full $U(1)^s$ gauge symmetry
being anomaly free
\foot{In fact, demanding anomaly freedom is a somewhat stronger
condition that the second Chern class constraint as the former
 amounts to requiring equality of the corresponding
differential four forms viewed as elements of a free module, i.e.
not taking into account cohomology relations. It would be interesting
to see if sense can be made of linear sigma models in which
only the latter, weaker condition is satisfied.}.
 In particular, if we denote the $U(1)^s$ gauge charges of
the right moving fermions $(\pi,\psi_1,...,\psi_5)$ by
$(q_0^{(k)},q_1^{(k)},...,q_5^{(k)})$ and the left moving fermions
$(\gamma,\lambda_1,...,\lambda_n)$ by
$(\q_0^{(k)},\q_1^{(k)},...,\q_n^{(k)})$ with $k = 1,...,s$, then the
conditions
generalizing \econe\ and \ectwo\ are
\eqn\econeg{ \q_0^{(k)} = \sum_{i=1}^5 q_i^{(k)}\  {\rm and}\
q_0^{(k)} = -\sum_{a=1}^n \q_a^{(k)} }
\eqn\ectwog{\sum_{i = 0}^5 q_i^{(j)} q_i^{(k)} = \sum_{a=0}^n \q_i^{(j)}
\q_i^{(k)} .}
With one of the $U(1)$ charges, say for $k = 1$, being those of the singular
model used in \esingularseqt, the other $U(1)^{s-1}$ charges give us the
data required to define the desingularization of the model.
Going back to \esuperpot\ we see that after choosing the left moving gauge
charges,
we must also modify our set of $n$ polynomials $F_a, a =1,...,n$ of the bosonic
fields to have multicharges
$(-q_0 - \q_a^{(1)}, -q_0 - \q_a^{(2)},...,-q_0 -\q_a^{(k)})$. Finally, we note
that in addition to these bosonic $U(1)^s$ gauge symmetries, we also have the
choice of imposing some number of independent fermionic gauge
symmetries.

If we impose $m$ such symmetries, we must introduce $m(n+1)$ homogeneous
polynomials $E^a_{(j)}, a=0,...,n; j = 1,...,m$ where
$E^a_{(j)}$ has $U(1)^s$ charges $(\q_a^{(1)},...,\q_a^{(s)})$, for all $j$.
These symmetries act on the fermionic chiral multiplets as
\eqn\efsymdef{\Gamma\to \Gamma +2 E^0_{(j)}(\Phi)\Omega^j,\quad \Lambda^a\to
\Lambda^a + 2E^a_{(j)}(\Phi)\Omega^j}
where the $\Omega^j$ are fermionic chiral superfields which are neutral under
the $U(1)^n$ gauge symmetry. Invariance of \esuperpot\ requires
\eqn\einv{E^0_{(j)}F_0 + P \sum_{a=1}^n E^a_{(j)} F_a = 0.}
To make the kinetic terms invariant, we need to introduce, for every
fermionic gauge symmetry, an unconstrained complex fermionic fermionic
superfield $\Sigma^j$, which transforms as $\Sigma^j\to\Sigma^j+\Omega^j$.
The detailed form of the action for the $\Sigma^j$ is explained in \DNotes.
For our purposes, it suffices to note that it leads to a term in the scalar
potential of the form
$$V_\sigma= \sum_j |\sigma_j|^2(
|E^a_{(j)}(\phi)|^2+|p|^2|E^0_{(j)}(\phi)|^2)$$
So long as, for each $j$, the quantity in parentheses is everywhere nonzero,
then the $\sigma_j$ are massive, and drop out of the infrared theory.

Now, if the $E^a_{(j)}$ for fixed but arbitrary $j$ with $a=1,...,n$ do not
simultaneously vanish on the Calabi--Yau, as well as the same being true
for the $F_a$ with $a=1,...,n$, then the same reasoning which led to
\esequence\ implies that the massless left moving fermions couple
to the vector bundle $V$ which is  the cohomology
of the sequence

\eqn\esequencetwo{0 \rightarrow \oplus^m {\cal O} {\buildrel
\otimes E^a_{(j)}\over\longrightarrow}
\oplus_{a=1}^n {\cal O}(\q_i^{(1)},\q_i^{(2)},...,\q_i^{(k)}) {\buildrel
\otimes F_a\over\longrightarrow}
 {\cal O}(\q_0^{(1)},\q_0^{(2)},...,\q_0^{(k)}) \rightarrow 0}
over the resolved space (where in this expression $a=1,...,n$).

If this condition can not be met on the $E$'s and $F$'s, $V$ may not
be a bundle on the resolved space. Depending on how severely the vanishing
conditions on these maps is violated, the model may still make perturbative
sense, as we shall see.

Notice that in general there isn't a unique set of charges $\q_i$ satisfying
\econeg\ and \ectwog. Thus, a given singular model may admit {\it many}\/
desingularizations. We emphasize that the well known and physically
important statement in $(2,2)$ theories \rAGM---that there can be distinct
ways of resolving the singularities of a given Calabi--Yau space refers
to the distinct maximal triangulations of the point set ${\cal P}$, for
a fixed and identifiable set of gauge charges.
Geometrically, this ambiguity corresponds to
the freedom of flopping certain rational curves in a given
Calabi--Yau yielding topological distinct cousins. In the $(0,2)$ setting we
see even
additional freedom in resolving the model as there are desingularizations
whose gauge charges differ as well.

We can summarize the algorithm for resolving a $(0,2)$ model on
a Calabi--Yau hypersurface $M$ in $\wp{4}{q_1,...,q_5}$ as follows:

${\bullet}$ Using the machinery of toric geometry, find the charges of the
$h^{11} + 5$ chiral superfields under the full $U(1)^{h^{11}}$ gauge symmetry
required to desingularize the base space $M$.

${\bullet}$ For a chosen rank of the gauge bundle, find a set of charges of
the
left moving fermions under this $U(1)^{h^{11}}$ gauge symmetry
meeting the anomaly cancelation conditions \econeg\ \ectwog.

${\bullet}$ Choose $m$ and a set of $F_a$ and $E^a_{(j)}$ $a=1,...,n$,
$j=1,...,m$
meeting the conditions given above.

Our discussion makes it apparent that there are a number of new features
that arise in the more general $(0,2)$ setting.
 From our perspective the two most prominent are:
(1) it is often the case that upon resolution of singularities in the base
Calabi--Yau the $F_a$'s can not be chosen to avoid them simultaneously
vanishing. Nonetheless, it can be arranged in many cases for the
vacuum field configuration space to remain compact thus giving rise
to an apparently well defined theory in which the left moving fermions
couple to a {\it sheaf}\/ rather than a bundle. Hence, this appears
to greatly broaden the geometrical setting for $(0,2)$ models.
(2) A given
$(0,2)$ model defined on  a singular Calabi--Yau space in the manner
of  \refs{\Witten,\DK,\DNotes} can admit distinct desingularizations. These, of
course,
are in addition to the known distinct desingularizations of the base
Calabi--Yau which played a prominent role in \refs{\Witten,\rAGM} and are of a
rather different character as the charges of the vacuum bundle differ and
hence appear to lead to a multi critical phenomenon. We give an example of
this below and will explore this aspect more fully elsewhere.

\newsec{Examples}

\subsec{Two Distinct (0,2) Resolutions}

For our first example, we consider $M$, a 12$^{th}$ order hypersurface in
$\wp{4}_{1,1,2,2,6}$. The (2,2) compactification on this manifold was
studied
using the techniques of mirror symmetry in \refs{\CdFKM,\HKTY}.

$M$ has a $\BZ_2$ orbifold singularity where $\phi_1=\phi_2=0$. To resolve
the singularity, we introduce another scalar, $\chi$, neutral under the
original $U(1)$, and a second $U(1)$ under which $\chi$ is  charged.

One (0,2) model that we could build on $M$ is, of course, a holomorphic
deformation of the tangent bundle. In that case, the $\lambda$s have exactly
the same charges as the $\phi$s, and the bundle $V_1$ is the cohomology of the
monad
\eqn\etandef{0\to\CO\oplus\CO
{\buildrel\otimes E^a_A(\phi)\over\longrightarrow}
\CO(1,-2)\oplus\CO(0,1)^{\oplus2}\oplus\CO(1,0)^{\oplus2}\oplus\CO(3,0)
{\buildrel\otimes F_a(\phi)\over \longrightarrow}\CO(6,0)\to0}

%\vfill

\def\newpar{\hfil\vadjust{\vskip\parskip}\break\indent}
\def\tablerule{\omit&
\multispan{6}{\tabskip=0pt\hrulefill}&\cr}
\def\tablepad{\omit&
height3pt&&&&&&\cr}
\leftline{\hsize=2.375in
\vbox{\offinterlineskip\tabskip=0pt\halign{
%\hskip.5in
\strut$#$\quad&
\vrule#&\quad\hfil $#$ \hfil\quad &\vrule #&\quad
\hfil $#$ \quad&\vrule #&\quad
\hfil $#$ \quad&\vrule #\cr
&\omit&\hbox{Field}&\omit&Q_1&\omit&Q_2&\omit\cr
\tablerule\tablepad
&& \phi_{1,2} &&0&&1&\cr
\tablepad\tablerule\tablepad
&& \phi_{3,4} &&1&&0&\cr
\tablepad\tablerule\tablepad
&& \phi_{5} &&3&&0&\cr
\tablepad\tablerule\tablepad
&& \chi&&1&&-2&\cr
\tablepad\tablerule\tablepad
&&p&&-6&&0&\cr
\tablepad\tablerule\tablepad\tablerule\tablepad
&&\lambda^{1,2}&&1&&-1&\cr
\tablepad\tablerule\tablepad
&&\lambda^{3}&&0&&2&\cr
\tablepad\tablerule\tablepad
&&\lambda^{4}&&1&&0&\cr
\tablepad\tablerule\tablepad
&&\lambda^{5}&&3&&0&\cr
\tablepad\tablerule\tablepad
&&\gamma&&-6&&0&\cr
\tablepad\tablerule
\noalign{\bigskip}
\noalign{\rightskip=30pt\noindent{\bf Table 1:}
$U(1)$ charges of the (bosonic and fermionic) fields for the resolved
model. The original charge is $Q=2Q_1+Q_2$.}
 }}}
\vskip-3.75in\hangindent=2.375in\hangafter=-18
For the tangent bundle,
$E^a_{1,2}=q_{1,2}^a\phi^a$ (no sum on $a$), and $F_a(\phi)=\pd{G}{\phi^a}$.
The quasihomogeneity of $G(\phi)$ implies
\eqn\eEF{ E_1^a(\phi)F_a(\phi)= 6 G(\phi),\quad E_2^a(\phi)F_a(\phi)=0} The
general bundle $V_1$ is defined by any set of $E$s and $F$s in \etandef\ which
satisfy \eEF.\newpar
 What is more interesting is that we can build another  rank three bundle,
$V_2$,
which is the cohomology of the monad
\eqn\eanotherV{\eqalign{0\to\CO{\to}
&\CO(1,-1)^{\oplus2}\oplus\CO(0,2)\oplus\cr
&\oplus\CO(1,0)\oplus\CO(3,0)\to\CO(6,0)\to0}}
\newpar
This is, on the resolved manifold, a completely different bundle than $V_1$,
and the (0,2) \lsm\ is very different. Rather than two fermionic gauge
symmetries (two $\Sigma$ multiplets) there is now only one. Correspondingly,
there are only five $\lambda$s instead of six. And, of course, the weights of
the $E^a$ and the $F_a$ are different
\foot{We are implicitly assuming the existence
of suitable $E$'s and $F$'s for this model, which in practice, can
be tedious to find explicitly.}.
Nevertheless, $V_1$ and $V_2$ become
isomorphic over the singular locus where we blow down the orbifold singularity.

This means, of course, that the net number of generation in the two models
must be the same. But, in fact, one can study the number of generations and
antigenerations separately for the two models. First, we consider built with
$V_1$, the deformation of the tangent bundle. We write the monad \etandef\ as
a pair of exact sequences
\eqn\epair{\eqalign{ 0\to\CO\oplus\CO\to
\CO(1,-2)\oplus\CO(0,1)^{\oplus2}&\oplus\CO(1,0)^{\oplus2}\oplus\CO(3,0)\to
\CE\to0\cr
0\to V_1&\to \CE\to \CO(6,0)\to 0\cr}}
Almost all of the line bundles in
\epair\ satisfy $h^i(\CO(n,m))=0$, $i>0$. The exceptions are $\CO$, which has $h^0(\CO)=h^3(\CO)=1$, and  $\CO(1,-2)$,
which has $h^0(\CO(1,-2))=1$ and $h^1(\CO(1,-2))=2$. The number of
holomorphic sections of the other line bundles can be determined by counting
monomials or by computing the index:
\eqn\eindex{{\rm Ind}\overline{\partial}_{\CO(n,m)}={n(13+2n^2)\over
3}+m(2+n^2)} Tracing through the long exact sequences in cohomology associated
to \epair, one finds that $h^1(V_1)=128$, and $h^2(V_1)=2$. These give,
respectively, the number of $\bf 27$s and $\overline{\bf 27}$s. One notes that
this is exactly the same number as in the (2,2) model. It might have happened,
that as one deformed away from (2,2), the extra $\bf 27$s and $\overline{\bf
27}$s paired up and became massive. This does not happen.

We can perform the same cohomological calculation for the bundle $V_2$,
defined by \eanotherV. Few of the details change. Again, only one of the line
bundles involved has nonvanishing higher cohomology groups. In this case, it
is $\CO(0,2)$ which has $h^0(\CO(0,2))=3$ and $h^2(\CO(0,2))=1$. Again,
tracing through the long exact sequences in cohomology, we find
$h^1(V_2)=128$, and $h^2(V_2)=2$.

For a little more insight, we can compute the spectrum of these two
theories at \LG. In fact, of course, at the \LG\ point, they are
isomorphic, so there is really only one calculation to do. $\gamma$,
$\lambda_6$ and $\chi$ are massive, , and after some rescaling, the \LG\
superpotential is
$$\CW=\int d\theta \sum_{a=1}^5\Lambda^aF_a(\Phi)$$
One finds 126 generations in the untwisted sector, realized as twelfth
order polynomials modulo the ideal generated by the $F_a$. There are two
more generations whose $16_{-1/2}$ components comes from the $k=12$ twisted
sector,
$\phi_0^{3,4}\ket{k=12}$. The ($16_{1/2}$ components of the) two
antigenerations are the ground states of the $k=6$ and $k=14$ sectors.

In this example, then we have seen two distinct resolutions of the vector
bundle when one resolves the manifold.  In our next example, we will see
that what one obtains on the resolved manifold need not even be a vector
bundle.

\subsec{Protuberances and Reflexive Sheaves}
Consider the manifold $M$, an octic hypersurface in
$\wp{4}_{1,1,2,2,2}$. The (2,2) compactification on this manifold was studied
using the techniques of mirror symmetry in \refs{\CdFKM,\HKTY}.
Instead of the tangent
bundle, we consider a rank three  vector bundle, $V$, defined as the kernel in
the exact sequence
\eqn\eVdeftwo{0\to V\to\CO^{\oplus 2}(1)\oplus\CO(2)\oplus\CO(5){\buildrel
\otimes F_a\over \longrightarrow}\CO(9)\to 0}
That is, in addition to the scalars $\phi_i$, with charges (1,1,2,2,2), and
the fermion $\gamma$, with charge $-8$, we have fermions $\lambda^a$, with
charges (1,1,2,5), and a scalar, $p$, with charge $-9$.

$M$ has a $\BZ_2$ orbifold singularity where $\phi_1=\phi_2=0$. To resolve
the singularity, we introduce another scalar, $\chi$, neutral under the
original $U(1)$, and a second $U(1)$ under which $\chi$ is  charged. The net
effect is to blow up the singularity, replacing each point on the orbifold
locus by a
$\cp{2}$.

%\vfill

\def\tablerule{\omit&
\multispan{6}{\tabskip=0pt\hrulefill}&\cr}
\def\tablepad{\omit&
height3pt&&&&&&\cr}
\leftline{\hsize=2.375in
\vbox{\offinterlineskip\tabskip=0pt\halign{
%\hskip.5in
\strut$#$\quad&
\vrule#&\quad\hfil $#$ \hfil\quad &\vrule #&\quad
\hfil $#$ \quad&\vrule #&\quad
\hfil $#$ \quad&\vrule #\cr
&\omit&\hbox{Field}&\omit&Q_1&\omit&Q_2&\omit\cr
\tablerule\tablepad
&& \phi_{1,2} &&0&&1&\cr
\tablepad\tablerule\tablepad
&& \phi_{3,4,5} &&1&&0&\cr
\tablepad\tablerule\tablepad
&& \chi&&1&&-2&\cr
\tablepad\tablerule\tablepad
&&p&&-5&&1&\cr
\tablepad\tablerule\tablepad\tablerule\tablepad
&&\lambda^{1,2}&&0&&1&\cr
\tablepad\tablerule\tablepad
&&\lambda^{3}&&2&&-2&\cr
\tablepad\tablerule\tablepad
&&\lambda^{4}&&3&&-1&\cr
\tablepad\tablerule\tablepad
&&\gamma&&-4&&0&\cr
\tablepad\tablerule
\noalign{\bigskip}
\noalign{\rightskip=30pt\noindent{\bf Table 2:}
$U(1)$ charges of the (bosonic and fermionic) fields for the resolved
model. The original charge is $Q=2Q_1+Q_2$.}
 }}}
\vskip-3.25in\hangindent=2.375in\hangafter=-17
The charge assignments of the fields under the two $U(1)$s are given in table
2. We have chosen a basis for the $U(1)$s such that the original $U(1)$ is
given by
$$Q=2Q_1+Q_2$$
The D-terms which follow from these charge assignments are
\eqn\eDtermstwo{\eqalign{
|\phi_3|^2+|\phi_4|^2+|\phi_5|^2+|\chi^2|-5|p|^2&=r_1\cr
|\phi_1|^2+|\phi_2|^2-2|\chi^2|+|p|^2&=r_2\cr
}}
\newpar
In the \cy\ phase, for $r_1,r_2 \gg0$, we expect that these D-terms, together
with the superpotential terms, force $\vev{p}=0$ (this turns out to be not
quite correct here, as we shall see shortly). Each point on the erstwhile
singularity $\phi_1=\phi_2=0$ is replaced by a $\cp{1}$, given by $\chi=0$,
$|\phi_1|^2+|\phi_2|^2=r_2$.\newpar
 In similar fashion, we expect that \eVdeftwo\ is
replaced by
\eqn\eVdeftworesol{
0\to\tilde V\to\CO^{\oplus 2}(0,1)\oplus\CO(2,-2)\oplus\CO(3,-1){\buildrel
\otimes F_a\over \longrightarrow}\CO(5,-1)\to 0}

The new feature about this sequence \eVdeftworesol\ is that
it is not exact! The problem is that
there are points on the resolved manifold $\tilde M$ where all of the
$F_a$'s vanish simultaneously. To see this, let us write the general form of
the $F_a$, denoting their charges by superscripts.
$$\eqalign{F_1^{(5,-2)}&=\chi \tilde F_1^{(4,0)}\cr
F_2^{(5,-2)}&=\chi \tilde F_2^{(4,0)}\cr
F_3^{(3,1)}&=\phi_1 f^{(3,0)}+ \phi_2 g^{(3,0)}\cr
F_4^{(2,0)}&\cr}$$
Note that setting $\chi=0$ {\it automatically}\/ guarantees $F_1=F_2=0$. To
satisfy the other two equations, we can set to zero all terms in $F_{3,4}$
which contain $\chi$. Then $F_4$ is a quadric in $\phi_3,\phi_4,\phi_5$,
and
$$
F_3=\phi_1 f +\phi_2 g$$
with $f,g$ being
cubics in $\phi_3,\phi_4,\phi_5$. For each solution to $F_4=0$, we
get a linear equation for $\phi_1,\phi_2$ from setting $F_3=0$.

Since, on setting $\chi=0$, the equation for $\tilde M$ is a quartic in
$\phi_3,\phi_4,\phi_5$, there are, all in all,
$8=4\times2$ points on $\tilde M$ where all of the $F$s vanish, all of
them located on the locus $\chi=0$. At these points, the sequence
\eVdeftworesol\ fails to be exact, because the last map isn't onto. To get an
exact sequence, we need include the cokernel of this map:
\eqn\eVdeftworesollong{
0\to\tilde V\to\CO^{\oplus 2}(0,1)\oplus\CO(2,-2)\oplus\CO(3,-1){\buildrel
\otimes F_a\over \longrightarrow}\CO(5,-1)\to \CO_S(5,-1)\to 0}
where the skyscraper sheaf $\CO_S(5,-1)$ is the restriction of the line
bundle $\CO(5,-1)$ to the set $S=\{\chi=F_3=F_4=0\}\subset \tilde M$.

The $\tilde V$ we obtain in this way, however, is not a vector bundle! It
fails to be locally free precisely at the points of $S$. If we remove those
points, $\tilde V$ is a vector bundle over $\tilde M\setminus S$. Even
over those ``bad'' points,
$\tilde V$ has the property of being isomorphic to its double dual
$(\tilde V)^{**}$.  Any sheaf which is isomorphic to its double dual is
called {\it reflexive}; such sheaves are
locally free, except for a ``singularity set" of complex codimension 3.
So in our low-energy nonlinear \sm, the left-moving
fermions couple to $\tilde V$, a rank three reflexive sheaf over $\tilde M$.

This is not all that surprising. We know that strings are perfectly happy
propagating on spaces which are not quite manifolds but, rather, have
certain suitably-mild singularities ({\it e.g.}~orbifolds). Here we see that
the left-movers can happily couple to objects which are not quite vector
bundles, but rather have certain suitably-mild singularities ({\it
e.g.}~reflexive sheaves).

Before we proceed to calculate the effect of this modification, let us
see how exactly this reflexive sheaf is realized in the \lsm?  The answer is
very simple. Away from the points in
$S$, the
\lsm\ just gives the same solution for $\tilde M$ that one expects from
geometry. However, at those points, where all the
$F$s vanish, $p$ is no longer forced to be zero. Instead, from the D-terms
\eDtermstwo, one finds that there is a $\cp{1}$ sitting over each
point in $S$.  The \lsm\ solution looks like the manifold $\tilde M$,
with 8 $\cp{1}$s glued on. These {\it protuberances}\/ are similar to the {\it
exoflops}\/ found in \rAGM\ in certain phases of (2,2) models. Here we would
like to interpret them as the
\lsm's way of coping with the skyscraper sheaf which makes \eVdeftworesol\
fail to be exact.

Let us compute the Chern classes of $\tilde V$, defined by
\eVdeftworesollong. Since, $h^{1,1}=2$, we have two fundamental cohomology
classes, $\eta_1, \eta_2$, which are, respectively, the first Chern classes
of the hyperplane bundles $\CO(1,0)$ and $\CO(0,1)$. The standard calculation
of the Stanley--Reisner ideal \BatQcoh\
for this manifold yields
\eqn\eSRtwo{
\eta_1^3=8,\qquad\eta_1^2\eta_2=4,\qquad\eta_2^2=0
}
To compute the Chern classes of $\CO_S(5,-1)$, we use the Koszul
complex\foot{For a complete intersection of divisors $D_i$, the Koszul
complex is
$$0\to \dots\bigoplus_{i<j<k}\CO(-D_i-D_j-D_k)\to
\bigoplus_{i<j}\CO(-D_i-D_j)\to \bigoplus_i \CO(-D_i)
\to\CO\to\CO_S\to 0$$}, which give  a locally free resolution of
$L_S$, for any line bundle $L$. Since $S$ is the complete intersection
${\chi=F_3=F_4=0}$, we have
\eqn\ekoszul{\eqalign{0\to L(-6,1)\to &L(-5,-1)\oplus L(-4,1)\oplus
L(-3,2)\cr&\to L(-3,-1)\oplus L(-2,0)\oplus L(-1,2)\to L\to L_S\to 0\cr}}
The total Chern class of $L_S$ is
\eqn\ecsky{c(L_S)= { c(L)c(L(-5,-1))c(L(-4,1))c(L(-3,2))\over
c(L(-3,-1))c(L(-2,0))c(-1,2))c(L(-6,1))}=1+12 \eta_1^3-20\eta_1^2\eta_2}
{}From \eSRtwo, we find $c_3(L_S)=16$.

Now we can compute the Chern classes of $\tilde V$.
\eqn\ectildeV{c(\tilde
V)=1+6\eta_1^2+2\eta_1\eta_2-18\eta_1^3-12\eta_1^2\eta_2}
So
\eqn\enumgentwo{c_3(\tilde V)=-192}
which predicts that the net number of generations is 96. Note that, without
the
correction due to
$c_3(\CO_S(-5,1))$, we would have gotten $c_3(\tilde V)=-208$.
Actually, with a little more work, we can do better than this and compute the
number of generations and anti-generations separately. From
\eVdeftworesollong, we can derive a long exact sequence in cohomology:
\eqn\elongvtilde{\eqalign{0\to
\H{0}{\CO(0,1)}^{\oplus2}\oplus& \H{0}{\CO(2,-2)}\oplus\H{0}{\CO(3,-1)}
\to\H{0}{\CO(5,-1)}\to\cr&\to\H{1}{\tilde V}
\to\H{1}{\CO(2,-2)}\to\H{0}{\CO_S(5,-1)}\to\H{2}{\tilde V}\to0\cr}}
The facts needed to derive this sequence are:
i) The higher cohomology groups for all of the line bundles (and the
skyscraper sheaf) which appear in \eVdeftworesollong\ vanish {\it except}\/
for $\H{1}{\CO(2,-2)}$, which is 6-dimensional and ii) the restriction of
$\H{0}{\CO(5,-1)}$ to the set $S$ vanishes.

The key point now is to study the map
$\H{1}{\CO(2,-2)}{\buildrel\alpha\over\to}\H{0}{\CO_S(5,-1)}$. On the overlap
of patches $U_{\{\phi_1\neq0\}}\cap U_{\{\phi_2\neq0\}}$,
$\H{1}{\CO(2,-2)}$ has representatives of the form
$${P_2(\phi_3,\phi_4,\phi_5)\over \phi_1\phi_2}$$
for a quadric polynomial $P_2$. (As announced, $h^1(\CO(2,-2))=6$.) The
restriction to the set
$S$ annihilates one of these quadric polynomial (which we called $F_4$ above),
so the cokernel of the map $\alpha$ has dimension $8-(6-1)=3$. So we conclude
that the number of antigenerations is $h^2(\tilde V)=3$ and, either using
the index \enumgentwo, or the explicit counts of holomorphic sections of the
line bundles appearing in \elongvtilde\ ($h^0(\CO(0,1))=2$,
$h^0(\CO(2,-2))=6$, $h^0(\CO(3,-1))=30$, $h^0(\CO(5,-1))=138$), we arrive a
the number of generations, $h^1(\tilde V)=99$.

Now let us turn to the \LG\ phase, and compute the massless spectrum at the
\LG\ point. The generations, ${\bf 27}$s of $E_6$, are obtained as follows.
(We state the results for the right-handed ${\bf 16}_{-1/2}$s of
$SO(10)\times U(1)$, that is, for the states with $({\bi q},\overline{\bi
q})=(-1/2,-1/2)$ arising from $k$-even sectors. The ${\bf 10}_{1}$ and ${\bf
1}_{-2}$ are obtained from the neighboring $k$-odd sectors.)

There are 98 generations obtained as nonic polynomials acting on the
untwisted vacuum, $P_9(\phi)\ket{k=0}$. These  nonic polynomials are taken
modulo the ideal generated by $W(\phi), F_a(\phi)$. There is also one more
generation, coming from the $k=10$ twisted sector, which has the form
$\lambda^3_{-1/9}\ket{k=9}$.

The 3 antigenerations ($({\bi q},\overline{\bi q})=(1/2,-1/2)$) also come from
the $k=10$ sector, and have the form
$\phi^{3,4,5}_{-1/9}\ket{k=10}$. All in all, there are 99 generations and 3
antigenerations, exactly as predicted by the geometrical analysis above.

One readily check that the $\BZ_{18}$ discrete R-symmetry which stems from
the quantum symmetry of the LG theory is nonanomalous. Recall that there
are possible anomalies due to $E_6$, $E_8$ and gravitational instantons.
The corresponding anomaly coefficients, $A_{1,2,3}$, must satisfy
\eqn\eanomcoefs{\eqalign{A_1&=A_2 \quad {\rm mod}\ 2 m r\cr
24 A_1=24A_2&=A_3  \quad {\rm mod}\ 2 m r\cr}}
where, here, $m=9$ and $r=3$.

We compute $A_1$ by taking a trace in the right-moving Ramond sector
(spacetime fermions) of tensor product of the ``internal" (0,2) SCFT and the
CFT consisting of 10 free left-moving Majorana--Weyl fermions, which carry the
$E_6$ gauge degrees of freedom:
$$A_1= Tr_R\left((kr-2{\bi q}) {{\bi q}^2\over 2r} (-1)^{F_R}\right)\quad
{\rm mod}\ 2mr$$ The charge $(kr-2{\bi q})$ act homogeneously on $E_6$
representations. The factor of ${{\bi q}^2\over 2r}$, being the square of an
$E_6$ generator, when traced gives the index of the corresponding $E_6$
representation. Also, we get equal contributions from each right-handed
fermion, and the left-handed charge-conjugate state, so, up to a factor of 2,
we can count the contribution only of right-handed fermions, weighted by the
index of the representation:
$$\eqalign{A_1&=2[(-3) c({\bf 78}) + 98(1)c({\bf 27})+(31)c({\bf
27})+3(29)c({\bf\overline{27}})]\cr
&=-18\quad{\rm mod}\ 54\cr}$$
where $c({\bf 78})=12$, $c({\bf 27})=c({\bf\overline{27}})=3$
$A_2$ receives contributions only from the gluinos of the second $E_8$, and
so is
$$A_2=-60r = -18 \quad{\rm mod}\ 54$$
The calculation of the gravitational contribution to the anomaly is the most
involved, because it receives contributions from all fermions, including the
gauge singlets.

The right-handed gauge singlets arise as follows: there are 318 which arise
from the ``untwisted" $k=1$ sector. These have the general form of an
oscillator mode of a fermion ($\lambda^a$ or $\gamma$) times a polynomial
of the appropriate degree in the $\phi$s.
19 singlets come from the $k=3$ sector,  and have the form of a quartic
polynomial in $\phi$ acting on the ground state.  2 come from the $k=5$
sector and have the form
$\bar\lambda^4_{-1/9}\phi^{1,2}_{-5/18}\ket{k=5}$, and 6 come from the
$k=11$ sector and have the form
$\lambda^{1,2}_{-1/9}\bar\phi^{1,2}_{-7/18}\ket{k=11}$ and
$\gamma_{-1/9}\bar\phi^{1,2}_{-7/18}\ket{k=11}$. There are also singlets
from the $k=9$ sector, but since they don't contribute to the anomaly, we
won't write them down.

Including by hand the contributions of the gluino, the dilatino and the
gluinos from the second $E_8$, the gravitational anomaly is
$$\eqalign{A_3&=-452r +Tr_R\left((kr-2{\bi q})  (-1)^{F_R}\right)\quad
{\rm mod}\ 2mr\cr
&=-452r+2[(-3)78+(98+31+3\cdot29)27+ 19(9)+ 2(15)+6(33)]\cr
&=0\quad {\rm mod}\ 2mr\cr
}$$
So the anomalies satisfy \eanomcoefs, and can be canceled by assigning a
suitable inhomogeneous transformation law to the axion.

\def\tablerule{\omit&
\multispan{6}{\tabskip=0pt\hrulefill}&\cr}
\def\tablepad{\omit&
height3pt&&&&&&\cr}
\leftline{\hsize=2.375in
\vbox{\offinterlineskip\tabskip=0pt\halign{
%\hskip.5in
\strut$#$\quad&
\vrule#&\quad\hfil $#$ \hfil\quad &\vrule #&\quad
\hfil $#$ \quad&\vrule #&\quad
\hfil $#$ \quad&\vrule #\cr
&\omit&\hbox{Field}&\omit&Q_1&\omit&Q_2&\omit\cr
\tablerule\tablepad
&& \phi_{1,2} &&0&&1&\cr
\tablepad\tablerule\tablepad
&& \phi_{3,4,5} &&1&&0&\cr
\tablepad\tablerule\tablepad
&& \chi&&1&&-2&\cr
\tablepad\tablerule\tablepad
&&p&&-5&&1&\cr
\tablepad\tablerule\tablepad\tablerule\tablepad
&&\lambda^{1,2}&&1&&-1&\cr
\tablepad\tablerule\tablepad
&&\lambda^{3}&&1&&0&\cr
\tablepad\tablerule\tablepad
&&\lambda^{4}&&3&&-1&\cr
\tablepad\tablerule\tablepad
&&\lambda^{4}&&-1&&2&\cr
\tablepad\tablerule\tablepad
&&\gamma&&-4&&0&\cr
\tablepad\tablerule
\noalign{\bigskip}
\noalign{\rightskip=30pt\noindent{\bf Table 3:}
$U(1)$ charges of the (bosonic
and fermionic) fields  for an unstable rank four bundle on the same manifold.}
 }}}
\vskip-3.5in\hangindent=2.375in\hangafter=-16
The resolved model that we have constructed has three antigenerations. It is
conceivable that, at \LG,  a flat direction exists where we turn on a ${\bf
27}$-$\overline{\bf 27}$ pair and break the gauge symmetry to $SO(10)$. At
first sight, there is an obvious candidate \lsm\ realization of this idea. We
start again with the same singular model as in the previous subsection. But
now, in the process of resolving the singularities, we ``borrow" a pair of
left-moving Majorana--Weyl fermions from the 10 free Majorana--Weyl fermions
which represent the gauge degrees of freedom. When we blow up, this pair
becomes an interacting Weyl fermion, an integral part of the ``internal" (0,2)
model.  The charge assignments are given in Table 3.
\newpar
The fifth fermion $\lambda^5$ is uncharged under the original $U(1)$,
$Q=2Q_1+Q_2$, and so is free along the singular locus. That is, along the
singular locus, the gauge group is $E_6$, as before. But, away from the
singular locus, the gauge group is broken to $SO(10)$.

The vacuum gauge bundle, $V'$,  on the resolved space $\tilde M$,
is the kernel in the exact sequence
\eqn\eVtildedefthree{0\to V'\to\CO(1,-1)^{\oplus
2}\oplus\CO(1,0)\oplus\CO(3,-1)\oplus\CO(-1,2)
{\buildrel
\otimes F_a\over \longrightarrow}\CO(5,-1)\to 0}
This time, there is no funny business; nowhere on $\tilde M$ do all of
the $F_a$ simultaneously vanish, so the sequence \eVtildedefthree\ really
is exact. $V'$ has rank four; along the singular locus, it is the direct
sum of a rank three bundle and a trivial line bundle. This is exactly what we
want to represent the breaking of $E_6$ to $SO(10)$ as we move away from the
singular locus.

Unfortunately, $V'$, as defined by \eVtildedefthree, is not stable. The
higher cohomology groups of all of the line bundles in \eVtildedefthree\
vanish {\it except}\/ for $\CO(-1,2)$, which has $h^3(\CO(-1,2))=1$ and
$h^2(\CO(-1,2))=3$. Tracing through the long exact sequence in cohomology
associated to \eVtildedefthree, one finds $h^1(V')=98$,
$h^2(V')=h^2(\CO(-1,2))=3$, and $h^3(V')=h^3(\CO(-1,2))=1$. The last is a
disaster. It means that $V'$ is not stable, as its dual bundle, $(V')^*$ has a
global section.

Physically, the postulated deformation, in which we turn on a
${\bf27}$-$\overline{\bf27}$ pair is not a flat direction.

In our next example, we will look at simple monad on a complete-intersection
Calabi-Yau manifold.

\subsec{A Monad Example}
\def\tablerule{\omit&
\multispan{6}{\tabskip=0pt\hrulefill}&\cr}
\def\tablepad{\omit&
height3pt&&&&&&\cr}
\leftline{\hsize=2.375in
\vbox{\offinterlineskip\tabskip=0pt\halign{
%\hskip.5in
\strut$#$\quad&
\vrule#&\quad\hfil $#$ \hfil\quad &\vrule #&\quad
\hfil $#$ \quad&\vrule #&\quad
\hfil $#$ \quad&\vrule #\cr
&\omit&\hbox{Field}&\omit&Q_1&\omit&Q_2&\omit\cr
\tablerule\tablepad
&& \phi_{1,2,3} &&0&&1&\cr
\tablepad\tablerule\tablepad
&& \phi_{4,5,6} &&1&&0&\cr
\tablepad\tablerule\tablepad
&& \chi&&1&&-3&\cr
\tablepad\tablerule\tablepad
&&p&&-3&&0&\cr
\tablepad\tablerule\tablepad\tablerule\tablepad
&&\lambda^{0}&&1&&-3&\cr
\tablepad\tablerule\tablepad
&&\lambda^{1,2,3}&&0&&1&\cr
\tablepad\tablerule\tablepad
&&\lambda^{4}&&2&&0&\cr
\tablepad\tablerule\tablepad
&&\gamma^{1,2}&&-2&&0&\cr
\tablepad\tablerule
\noalign{\bigskip}
\noalign{\rightskip=30pt\noindent{\bf Table 4:}
$U(1)$ charges of the (bosonic and fermionic) fields for the resolved
model. The original charge is $Q=3Q_1+Q_2$.}
 }}}
\vskip-3.25in\hangindent=2.375in\hangafter=-17
Consider the complete intersection of two sextics in
$\wp{5}_{1,1,1,3,3,3}$. This has a $\BZ_3$ orbifold singularity at
$\phi_1=\phi_2=\phi_3=0$. We torically blow up the singularity in the usual
way, which adds a second $U(1)$ and a field $\chi$. For the vector bundle, we
choose a bundle of rank 3\foot{This example also arose in conversations with
T.~M.~Chiang.}. On the original, unresolved manifold, this is a kernel,
$$0\to V\to \CO(1)^{\oplus3}\oplus\CO(6)\to \CO(9)\to 0$$
On the resolved
space, it is the cohomology of the monad
\eqn\egoodness{\eqalign{0\to
\CO\to\CO(1,-3)&\oplus\CO(0,1)^{\oplus3}\oplus\CO(2,0)\to\cr&\to\CO(3,0)\to0
\cr}}
The fields and their charges under the two $U(1)$s are listed in Table
4.\newpar
Working out the Stanley-Reisner ideal for this manifold, we find
\eqn\golly{\eta_1^3=36,\quad\eta_1^2\eta_2=12,\quad\eta_1\eta_2^2=4,\quad
\eta_2^3=0}
so $c_3(T)=-4\eta_1^3-3\eta_1^2\eta_2+9\eta_1\eta_2^2=-144$ and
$c_3(V)=-6\eta_1^3-3\eta_1^2\eta_2+9\eta_1\eta_2^2=-216$. Thus the net number
of generations is $108$.

We can, as in the previous examples, calculate $h^1(V)$ and $h^2(V)$
separately. Split the monad \egoodness\ into two short exact sheaf sequences,
\eqn\molly{\eqalign{
0\to\CO\to\CO(1,-3))\oplus&\CO(0,1)^{\oplus3}\oplus\CO(2,0)\to \CE\to0\cr
0\to&V\to\CE\to \CO(3,0)\to0\cr
}}
Of the line bundles in \molly, only $\CO$ and $\CO(1,-3)$ have nonzero higher
cohomology groups:
$$h^0(\CO)=h^0(\CO(1,-3))=h^3(\CO)=1,\quad h^2(\CO(1,-3))=3$$
$\H{2}{\CO(1,-3)}$ is generated by the cocycles
$\phi_{4,5,6}\over\phi_1\phi_2\phi_3$ on the triple overlap
$U_1\cap U_2\cap U_3$. The number of global sections of the other line
bundles is given by the index:
\eqn\richard{\eqalign{h^0(\CO(n,m))&={1\over12}\left(c_1(\CO(n,m))c_2(T)+2c_
1(\CO(n,m))^3\right)\cr
&=7n+3m+6n^3+6n^2m+2nm^2\cr}}
Putting all this together, we find
$$\H{1}{V}={\H{0}{\CO(3,0)}\over\H{0}{\CO(1,-3)}\oplus\H{0}{\CO(0,1)}^{\oplus3}
\oplus\H{0}{\CO(2,0)}}$$
which yields $h^1(V)=112$ and we find the
exact sequence
$$0\to \H{2}{\CO(1,-3)}\to\H{2}{V}\to\H{3}{\CO}\to0$$
which yields $h^2(V)=4$.

 This is readily compared with the spectrum at
Landau-Ginzburg. There are 112
$\bf 27$s of $E_6$ of the form $P_9(\phi)\ket{k=0}$ from the untwisted sector.
There are a total of four $\overline{\bf 27}$s from the twisted sectors. Two
are the ground states of the $k=6$ and $k=8$ twisted sectors; the remaining
two have the form $\phi_0^{4,5,6}\ket{k=6}$, modulo the $\bar Q^+
\bar\lambda_0^4\ket{k=6}$. So, indeed, the number of generations is unchanged
from the prediction at large radius.

For our last example, we construct a (0,2) model on K3

\subsec{A K3 example}
\def\tablerule{\omit&
\multispan{6}{\tabskip=0pt\hrulefill}&\cr}
\def\tablepad{\omit&
height3pt&&&&&&\cr}
\leftline{\hsize=2.375in
\vbox{\offinterlineskip\tabskip=0pt\halign{
%\hskip.5in
\strut$#$\quad&
\vrule#&\quad\hfil $#$ \hfil\quad &\vrule #&\quad
\hfil $#$ \quad&\vrule #&\quad
\hfil $#$ \quad&\vrule #\cr
&\omit&\hbox{Field}&\omit&Q_1&\omit&Q_2&\omit\cr
\tablerule\tablepad
&& \phi_{1,2} &&0&&1&\cr
\tablepad\tablerule\tablepad
&& \phi_{3,4,5} &&1&&0&\cr
\tablepad\tablerule\tablepad
&& \chi&&1&&-2&\cr
\tablepad\tablerule\tablepad
&&p&&-3&&1&\cr
\tablepad\tablerule\tablepad\tablerule\tablepad
&&\lambda^{1,2}&&0&&1&\cr
\tablepad\tablerule\tablepad
&&\lambda^{3}&&1&&-1&\cr
\tablepad\tablerule\tablepad
&&\lambda^{4}&&2&&-2&\cr
\tablepad\tablerule\tablepad
&&\gamma^{1,2}&&-2&&0&\cr
\tablepad\tablerule
\noalign{\bigskip}
\noalign{\rightskip=30pt\noindent{\bf Table 5:}
$U(1)$ charges of the (bosonic and fermionic) fields for the resolved
model. The original charge is $Q=2Q_1+Q_2$.}
 }}}
\vskip-3.25in\hangindent=2.375in\hangafter=-16
We consider a rank-3 bundle on K3, which we take to be the intersection of
two quartics in $\wp{4}_{1,1,2,2,2}$. Upon resolving the singularities, this
toric construction gives a two-dimensional slice through the
twenty-dimensional \Ka\ moduli space of K3. The bundle is constructed as a
kernel (no fermionic gauge symmetries) and, unlike the previous example,
remains a kernel when resolved. The charges of the fields for the resolved
manifold are listed in Table 5. The bundle $V$ is given by
\eqn\zerbina{\eqalign{0\to V\to
\CO(0,1)^{\oplus2}&\oplus\CO(1,-1)\oplus\CO(2,-2)\cr
\to&\CO(3,-1)\to0}}
The number of hypermultiplets is given by the index
\eqn\zippy{h^1(V)=-{\rm Ind}(\overline{\partial}_V)=c_2(V)-2r=18}
since we have set $c_2(V)=c_2(T)=24$ for K3, and $r(V)=3$.\newpar

We can compare this with the Landau-Ginzburg calculation. How this goes in
six dimensions may be a little unfamiliar, so let us pause to make a few
comments. As far as the left-moving degrees of freedom (which assemble the
gauge representations), the structure is the same as what we have seen in four
dimensions. However, the right-movers work a little differently. The fermions
in six-dimensional vector multiplets have $\bar q=\pm1$. As in the
four-dimensional case, half of them (those with $\bar q=-1$) come in a
completely canonical way from the $k=0,1,2$ sectors. The others come from the conjugate sectors (in
this case, $10-k$). The fermions in hypermultiplets have $\bar
q=0$, but again come in pairs, respectively in representations $R$ and
$\overline R$ of the gauge group, from conjugate twisted sectors of the LG
orbifold. In our case, we will simply list which sectors the $\bf27$s
(actually, just the ${\bf16}_{-1/2}$ components of the $\bf 27$) arise in;
the $\overline{\bf27}$s arise in the conjugate sectors.

One finds 14 $\bf27$s in the untwisted sector, realized as
$P_5(\phi)\ket{k=0}$. There are 3 $\bf27$s in the $k=4$ sector,
$\bar\phi^{3,4,5}_{-1/5}\ket{k=4}$, and one $\bf27$ in the $k=6$ sector,
$\lambda^4_{-1/5}\ket{k=6}$. All in all, this is $14+3+1=18$ hypermultiplets,
as expected.

\newsec{Prospects}

As we have seen, the \lsm\ allows one to define a wider class of (0,2)
theories than previously considered. Rather than having the left-moving
fermions couple to a stable holomorphic vector bundle, the \lsm\ is perfectly
happy having the left-moving fermions coupling to a stable torsion-free (or
reflexive) sheaf. Such a sheaf is a vector bundle outside of a ``singularity
set" of complex codimension 2 (in the case of a torsion-free sheaf) or 3 (in
the case of a reflexive sheaf).

The linear $\sigma$-model realizes the singularity set as a protuberance, that
is, as an additional branch of zeroes of the scalar potential, glued in at
the location of the singularity set on $M$. In the first example
in section 4.2, the singularity
set consisted of 8 points, and over each point, the protuberance was a
$\cp{1}$. The key point is that the protuberance is {\it compact}, so there
is no divergence of the linear $\sigma$-model, and hence, presumably, no
divergence of the superconformal field theory to which is flows in the
infrared.

Since there was no divergence, we can presume that no interesting
nonperturbative physics is associated with this type of protuberance. In
particular, there is no signal of new massless states appearing for those
values of the moduli.

There are, however, situations where new nonperturbative physics is dictated.
First, in some cases, it can happen that, when all of the $F_a(\phi)$ vanish
at some point in $M$, the new branch of the space of zeroes of the scalar
potential is {\it noncompact}. This is the (0,2) analogue of the (complex
structure side of) the conifold, in which the polynomials defining the
manifold $M$ fail to be transverse.

Another possibility, when $V$ is defined as the cohomology of a monad, is that
at some point in the K\"ahler moduli space, the $E_a$s might all
simultaneously vanish. In this case, we again get a noncompactness of the
space of zeroes of the scalar potential, this time associated to the complex
scalar $\sigma$ in the $\Sigma$ multiplet. This is the (0,2) analogue of the
K\"ahler side of the conifold (the singularity in the K\"ahler moduli space at
$r=\theta=0$).

In both of these cases, the \lsm, and hence the superconformal field theory,
is singular. We hope to discuss the physics of these singularities in a
future work.\foot{As this manuscript was nearing completion, we
learned of work of Kachru, Seiberg and Silverstein \rKSS\ which addresses
this issue, in the case of (0,2) models obtained as deformations of
(2,2) models.}

%\foot{while this manuscript was in preparation, we learned
%that this issue, in the case of (0,2) models obtained as deformations of
%(2,2) models, would be addressed in a forthcoming work of Kachru, Seiberg
%and Silverstein.}

Some examples of the physics which can ensue are provided by the work of
Witten and collaborators on ``small instantons"
\refs{\smallinstantons,\DMW}.
This is a particular example of the first type
of singularity, where the
$F$s simultaneously vanish on some set of complex codimension 2 on
$M$. In the case of $M={\rm K3}$ (itself of complex dimension 2), this
singularity set is a collection of points, the locations of instantons
shrunken to zero size. The findings of  \refs{\smallinstantons,\DMW} can be
translated into the language
of torsion-free sheaves as follows: for each point in the singularity set
which contributes $n$ to $c_2(V)$ (we persist in calling the torsion-free
sheaf which couples to the left-movers ``$V$"), there arises a
nonperturbative $Sp(n)$ gauge symmetry in 6 dimensions.
When $M$ has complex dimension 3 ({\it i.e.}~a Calabi--Yau manifold), the
singularity set could be complex codimension 2 {\it or}\/ 3.

An exciting prospect is that at some of these points where nonperturbative
physics is required, our experience with $(2,2)$ models leads us
to suspect that there may well be additional branches of the $(0,2)$
moduli space that can be reached in a physically smooth manner. Some
such transitions would presumably involve  topology changing transitions
of the base Calabi-Yau manifolds together with nontrivial changes in
the bundle structure as well. We hope to report on such processes
shortly.

\bigskip
\centerline{\bf{Acknowledgments}}

The research of J.D.~is supported by NSF grant
PHY-9511632, the Robert A.~Welch Foundation and an Alfred P.~Sloan Foundation
Fellowship.  The research of B.R.G.~is supported by the
Alfred P. Sloan Foundation, a National Young
Investigator Award, and by the National Science Foundation.
The research of D.R.M. is supported by NSF grant DMS-9401447.

\listrefs
\end